\documentclass[twocolumn,aps,prl,showpacs]{revtex4}
\usepackage{graphicx}

\begin{document}

\author{Yaroslav Tserkovnyak and Arne Brataas}
\affiliation{Harvard University, Lyman Laboratory of Physics, Cambridge, 
Massachusetts
02138}
\author{Gerrit E. W. Bauer}
\affiliation{Department of Applied Physics and DIMES, Delft University of 
Technology,
2628 CJ Delft, The Netherlands}
\title{Enhanced Gilbert Damping in Thin Ferromagnetic Films}

\begin{abstract}
The precession of the magnetization of a ferromagnet is shown to transfer
spins into adjacent normal metal layers. This
\textquotedblleft pumping\textquotedblright
of spins slows down the precession corresponding
to an enhanced Gilbert damping constant in the Landau-Lifshitz equation. The
damping is expressed in terms of the scattering matrix of the
ferromagnetic layer, which is accessible to model and
first-principles calculations. Our estimates for permalloy thin films
explain the trends observed in recent experiments.
\end{abstract}

\date{\today }
\maketitle

The magnetization dynamics of a bulk ferromagnet is well described by the
phenomenological Landau-Lifshitz-Gilbert (LLG) equation \cite{Gilbert:pr55} 
\begin{equation}
\frac{d\mathbf{m}}{dt}=-\gamma \mathbf{m}\times \mathbf{H}_{\text{eff}
}+\alpha \mathbf{m}\times \frac{d\mathbf{m}}{dt}\,,  \label{llg}
\end{equation}
where $\mathbf{m}$ is the magnetization direction, $\gamma $ is the
gyromagnetic ratio, and $\mathbf{H}_{\text{eff}}$ is the effective magnetic
field including the external, demagnetization, and crystal anisotropy
fields. The second term on the right-hand side of Eq.~(\ref{llg}) was first
introduced by Gilbert \cite{Gilbert:pr55} and the dimensionless coefficient $
\alpha $ is called the Gilbert damping constant. For a constant $\mathbf{H}_{
\text{eff}}$ and $\alpha =0$, $\mathbf{m}$ precesses around the field vector
with frequency $\omega =\gamma H_{\text{eff}}$. When damping is switched on $
\alpha >0$, the precession spirals down to a time independent magnetization
along the field direction on a time scale of $1/\alpha \omega $. Study of $
\alpha $ in bulk metallic ferromagnets has drawn a significant interest over
several decades. Notwithstanding the large body of both experimental \cite
{Bhagat:prb74} and theoretical \cite{Korenman:prb72} work, the damping
mechanism in bulk ferromagnets is not yet fully understood.

The magnetization dynamics in thin magnetic films and microstructures is
technologically relevant for, \textit{e.g.}, magnetic recording applications
at high bit densities. Recent interest of the basic physics community in
this topic is motivated by the spin-current induced magnetization switching
in layered structures \cite{Sloncz:mmm96,Myers:sc99,Katine:prl00}. The
Gilbert damping constant was found to be $0.04<\alpha <0.22$ for Cu-Co
and Pt-Co \cite{Back:sc99,Myers:sc99}, which is considerably larger than the
bulk value $\alpha _{0}\approx 0.005$ in Co \cite
{Schreiber:ssc95,Katine:prl00}. Previous attempts to explain the additional
damping in magnetic multilayer systems involved an enhanced electron-magnon
scattering near the interface \cite{Berger:prb96} and other mechanisms \cite
{Wegrowe:prb00}, both in equilibrium and in the presence of a spin-polarized
current.

In this Letter we propose a novel mechanism for the Gilbert damping in
normal-metal--ferromagnet (\textit{N-F}) hybrids. According to Eq.~(\ref{llg}),
the precession of the magnetization direction $\mathbf{m}$ is caused by
the torque $\propto \mathbf{m}\times \mathbf{H}_{\text{eff}}$. This is
physically equivalent to a volume injection of what we call a
\textquotedblleft spin current\textquotedblright . The damping occurs when
the spin current is allowed to leak into a normal metal in contact with the
ferromagnet. Our mechanism is thus the inverse of the spin-current induced
magnetization switching: A spin current can exert a finite torque on the
ferromagnetic order parameter, and, \textit{vice versa}, a moving
magnetization vector loses torque by emitting a spin current. In other
words, the magnetization precession acts as a spin pump which transfers
angular momentum from the ferromagnet into the normal metal. This effect can
be mathematically formulated in terms of the dependence of the scattering
matrix of a ferromagnetic layer attached to normal metal leads on the
precession of $\mathbf{m}$, analogous to the parametric charge pumping in
nonmagnetic systems \cite{Brouwer:prb98}. The damping contribution is found
to obey the LLG phenomenology. Enhancement of the damping constant $\alpha
^{\prime }=\alpha -\alpha _{0}$ can be expressed in terms of the scattering
matrix at the Fermi energy of a ferromagnetic film in contact with normal
metal reservoirs, which can be readily obtained by model or first-principles
calculations. Our numerical estimates of $\alpha ^{\prime }$ compare well
with recent experimental results \cite{Mizukami:jjap01}. Earlier experiments
reported in Ref.~\cite{Heinrich:prl87} can also be understood by our
model \cite{Tserkovnyak:prep}.

\begin{figure}[tbp]
\includegraphics[scale=0.45]{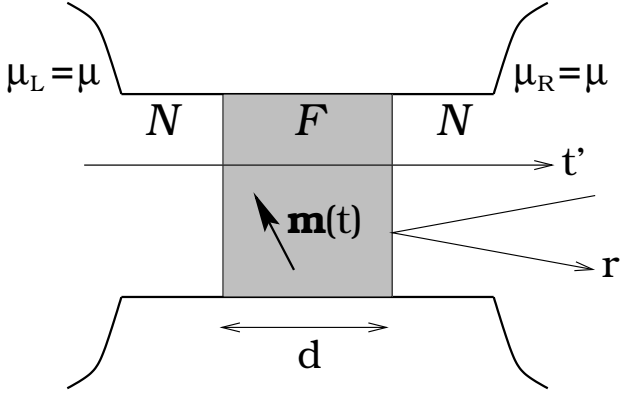}
\caption{Ferromagnetic film (\textit{F}) adjacent to two normal metal layers
(\textit{N}). The latter are viewed as reservoirs in common thermal
equilibrium. The reflection and transmission amplitudes $r$ and $t'$ shown
here govern the spin current pumped into the right lead.}
\label{f1}
\end{figure}

We consider a ferromagnetic film sandwiched between two paramagnetic layers
as shown in Fig.~\ref{f1}.
Spin pumping is governed by the
ferromagnetic film and the vicinity of the \textit{N-F}
interfaces.
The normal metal layers are, therefore, interpreted as
reservoirs attached to non-magnetic leads.
The quantity of interest is the $2\times 2$ current
matrix in spin space $\hat{I}=\hat{1}I_{c}/2-\hat{\text{\boldmath$\sigma $}}
\cdot \mathbf{I}_{s}e/\hbar $ for the charge ($I_{c}$) and spin flow ($
\mathbf{{I}_{s}}$) from the magnetic film into adjacent normal metal leads,
where $\hat{1}$ is the unit matrix and $\hat{\text{\boldmath$\sigma $}}$ the
vector of Pauli spin matrices.

When no voltages are applied and the external field is constant, the charge
current vanishes. Two contributions to the spin current $\mathbf{I}_{s}$
on either side of the ferromagnet may be distinguished, \textit{viz.}
$\mathbf{I}_{s}^{\text{pump}}$ and $\mathbf{I}_{s}^{(0)}$.
$\mathbf{I}_{s}^{\text{pump}}$ is the spin current pumped into the
normal metal to be discussed below,
whereas $\mathbf{I}_{s}^{(0)}$ is the current which flows
back into the ferromagnet. The latter is driven by the accumulated spins
in the normal metal and gives, \textit{e.g.}, rise to the spin-current induced 
magnetization switching \cite{Sloncz:mmm96,Myers:sc99,Katine:prl00}. Here, we
model the normal metal as an ideal sink for the spin current, such that a
spin accumulation does not build up. This approximation is valid when the spins
injected by $\mathbf{I}_{s}^{\text{pump}}$ decay and/or leave the interface
sufficiently fast, \textit{e.g.}, when the dimensionless conductance of the 
\textit{N-F} interface is smaller than $h/\tau_{\text{sf}}\delta$
\cite{Tserkovnyak:prep}. Here, $\tau_{\text{sf}}$ is the spin-flip
relaxation time and $\delta$ is the energy level spacing at the Fermi surface
of a normal metal film with a thickness which is the smaller one of the
geometrical film thickness and the spin-flip diffusion length.

The current $\hat{I}(t)$ pumped by the precession of the magnetization into
the right and left paramagnetic reservoirs, connected to the ferromagnet by
normal metal leads ($R$) and ($L$), may be calculated in an adiabatic
approximation since the period of precession $2\pi /\omega $ is typically
much larger than the relaxation times of the electronic degrees of freedom
of the system. The adiabatic charge-current response in nonmagnetic systems
by a scattering matrix which evolves under a time-dependent system parameter 
$X(t)$ has been derived in \cite{Buttiker:zpb94,Brouwer:prb98}. Adopting 
Brouwer's
notation \cite{Brouwer:prb98}, the generalization to the $2\times 2$
matrix current (directed into the normal metal lead $l=R$ or $L$) reads 
\begin{equation}
\hat{I}(t)^{\text{pump}}=e\frac{\partial \hat{n}(l)}{\partial X}\frac{dX(t)}{
dt}\,,  \label{Ip}
\end{equation}
where the matrix emissivity into the lead $l$ is 
\begin{equation}
\frac{\partial \hat{n}(l)}{\partial X}=\frac{1}{4\pi i}\sum_{mnl^{\prime }}
\frac{\partial \hat{s}_{mn,ll^{\prime }}}{\partial X}\hat{s}_{mn,ll^{\prime
}}^{\dagger }+\text{H.c.}  \label{em}
\end{equation}
and $\hat{s}$ is the $2\times 2$ scattering matrix of the ferromagnetic
insertion. $m$ and $n$ label the transverse modes at the Fermi energy in the
normal metal leads and $l^{\prime }=R,L$. Spin-flip scattering in the
contact is disregarded. $\hat{s}$ depends on the magnetization direction $
\mathbf{m}$ of the ferromagnet through the projection matrices $\hat{u}
^{\uparrow }=\left( \hat{1}+\hat{\text{\boldmath$\sigma $}}\cdot \mathbf{m}
\right) /2$ and $\hat{u}^{\downarrow }=\left( \hat{1}-\hat{\text{\boldmath$
\sigma $}}\cdot \mathbf{m}\right) /2$ (Ref.~\cite{Brataas:prl00}): $
\hat{s}_{mn,ll^{\prime }}=s_{mn,ll^{\prime }}^{\uparrow }\hat{u}^{\uparrow
}+s_{mn,ll^{\prime }}^{\downarrow }\hat{u}^{\downarrow }$. The spin current
pumped by the magnetization precession is obtained by identifying $
X(t)=\varphi (t)$, where $\varphi $ is the azimuthal angle of the
magnetization direction in the plane perpendicular to the precession axis.
The resulting current is traceless, $\hat{I}^{\text{ pump}}=-(e/\hbar )\hat{
\text{\boldmath$\sigma $}}\cdot \mathbf{I}_{s}^{\text{pump}}$, \textit{i.e.},
charge current indeed vanishes, and 
\begin{equation}
\mathbf{I}_{s}^{\text{pump}}=\frac{\hbar }{4\pi }\left( A_{r}\mathbf{m}
\times \frac{d\mathbf{m}}{dt}-A_{i}\frac{d\mathbf{m}}{dt}\right) \,,
\label{Is}
\end{equation}
where the interface parameters are 
\begin{eqnarray}
A_{r} &=&\frac{1}{2}\sum_{mn}\{\left\vert r_{mn}^{\uparrow
}-r_{mn}^{\downarrow }\right\vert ^{2}+\left\vert t_{mn}^{\prime \uparrow
}-t_{mn}^{\prime \downarrow }\right\vert ^{2}\}\,,  \label{Ar} \\
A_{i} &=&\mbox{Im}\sum_{mn}\{r_{mn}^{\uparrow }(r_{mn}^{\downarrow })^{\ast
}+t_{mn}^{\prime \uparrow }(t_{mn}^{\prime \downarrow })^{\ast }\}\,.
\label{Ai}
\end{eqnarray}
Here, $r_{mn}^{\uparrow }$ [$r_{mn}^{\downarrow }$] is the reflection
coefficient for spin-up [spin-down] electrons in the $l$th lead and $
t_{mn}^{\prime \uparrow }$ [$t_{mn}^{\prime \downarrow }$] is the
transmission coefficient for spin-up [spin-down] electrons into the $l$th
lead. (See Fig.~\ref{f1} for $l=R$.) Using unitarity of the scattering
matrix for each spin direction, we can summarize Eqs.~(\ref{Ar}) and (\ref
{Ai}) by $A_{r}+iA_{i}=g^{\uparrow \downarrow }-t^{\uparrow \downarrow }$,
where $g^{\sigma \sigma ^{\prime }}=\sum_{mn}\{\delta _{mn}-r_{mn}^{\sigma
}(r_{mn}^{\sigma ^{\prime }})^{\ast }\}$ is the (DC) conductance matrix \cite
{Brataas:prl00,Waintal:prb00} and $t^{\uparrow \downarrow
}=\sum_{nm}t_{mn}^{\prime \uparrow }(t_{mn}^{\prime \downarrow })^{\ast }$.
The spin current (\ref{Is}) trivially vanishes for the steady state, \textit{
i.e.}, when $d\mathbf{m}/dt=0$, and for unpolarized contacts $
s_{mn,ll^{\prime }}^{\uparrow }=s_{mn,ll^{\prime }}^{\downarrow }$.

Per revolution, the precession pumps an angular momentum into an adjacent
normal metal layer which is proportional to $A_{r}$, in the direction of the
(averaged) applied magnetic field, and decaying in time. At first sight, it
is astonishing that a pump can be operated by a single parameter varying in
time, whereas the ``peristaltic'' pumping of a charge current requires at
least two periodic parameters \cite{Brouwer:prb98}. However, there are
actually two periodic parameters (out of phase by $\pi/2$) hidden behind $
\varphi(t)$, \textit{viz.} the projections of the unit vector defined by $
\varphi$ in the plane perpendicular to the axis of precession.

By conservation of angular momentum, the spin torque on the ferromagnet
resulting from the spin pumping into the nonmagnetic leads gives an
additional term to the LLG equation (\ref{llg}). After including this term,
Eq.~(\ref{llg}) remains valid, but the gyromagnetic ratio and the damping
constant are renormalized: 
\begin{eqnarray}
\frac{1}{\gamma } &=&\frac{1}{\gamma _{0}}
\{1+g_{L}[A_{i}^{(L)}+A_{i}^{(R)}]/4\pi M\}\,,  \label{c1} \\
\alpha  &=&\frac{\gamma }{\gamma _{0}}\{\alpha
_{0}+g_{L}[A_{r}^{(L)}+A_{r}^{(R)}]/4\pi M\}\,.  \label{c2}
\end{eqnarray}
Here, $g_{L}$ is the Land\'{e} factor and $M$ is the total magnetic moment (in
units of $\mu _{B}$) of the ferromagnetic film; subscript $0$ denotes the
bulk values of $\gamma $ and $\alpha $; superscripts $(L)$ and $(R)$ denote
parameters evaluated on the left and right side of the \textit{F} layer,
respectively. Eqs.~(\ref{c1}) and (\ref{c2}) are the central result of this
paper.  $A_{r}$ and $A_{i}$ affect, \textit{e.g.}, ferromagnetic resonance
experiments as a shift of the resonance magnetic field via $
A_{i}^{(L)}+A_{i}^{(R)}$, whereas $A_{r}^{(L)}+A_{r}^{(R)}$ increases the
relative resonance linewidth.

From now on we focus on ferromagnetic films which are thicker than the
coherence length $\lambda _{\text{fc}}=\pi /(k_{\uparrow
}-k_{\downarrow })$, where $k_{\uparrow \downarrow }$ are the spin-up and
spin-down Fermi wavevectors, \textit{i.e.}, thicker than a few monolayers in
the case of transition metals. In this regime, spin-up and spin-down
electrons transmitted or scattered from one \textit{N-F} interface interfere
incoherently at the other interface,  $t^{\uparrow \downarrow }$ vanishes 
and the mixing conductance $g^{\uparrow \downarrow }$
is governed by the reflection coefficients of the isolated \textit{N-F}
interfaces.

$A_{i}=\mbox{Im}g^{\uparrow \downarrow }$ vanishes for ballistic and
diffusive contacts as well as nonmagnetic tunnel barriers \cite{Brataas:prl00}.
First-principles calculations find very small $A_{i}$ for
Cu-Co and Fe-Cr \cite{Xia:submitted}. It is, therefore, likely that $A_{i}$
may be disregarded in many systems. If $A_{i}$ does vanish on both sides of
the ferromagnetic film, it follows from Eqs.~(\ref{c1}) and (\ref{c2}) that
the resonance frequency is not modified $\gamma =\gamma _{0}$ and the
enhancement of the Gilbert damping is given by $\alpha ^{\prime
}=g_{L}[A_{r}^{(L)}+A_{r}^{(R)}]/4\pi M$.

The coefficient $A_{r}$ can be estimated by simple model calculations \cite
{Brataas:prl00}. For ballistic (point) contacts, $A_{r}^{B}=(1+p)g$ with the
polarization $p=(g^{\uparrow \uparrow }-g^{\downarrow \downarrow
})/(g^{\uparrow \uparrow }+g^{\downarrow \downarrow })$ and the average
conductance $g=(g^{\uparrow \uparrow }+g^{\downarrow \downarrow })/2$. For
diffusive \textit{N-F} hybrids, $A_{r}^{D}=g_{N}$,
the conductance of the normal metal part.
A nonmagnetic tunneling barrier between \textit{F} and \textit{N} suppresses
the spin current exponentially. The magnetization precession of a magnetic
insulator can also emit a spin current into a normal metal, since $
g^{\uparrow \downarrow }$ does not necessarily vanish because the phase
shifts of reflected spin-up and spin-down electrons at the interface may
differ \cite{Xia:submitted}.

\begin{figure}[ptb]
\includegraphics[scale=0.38]{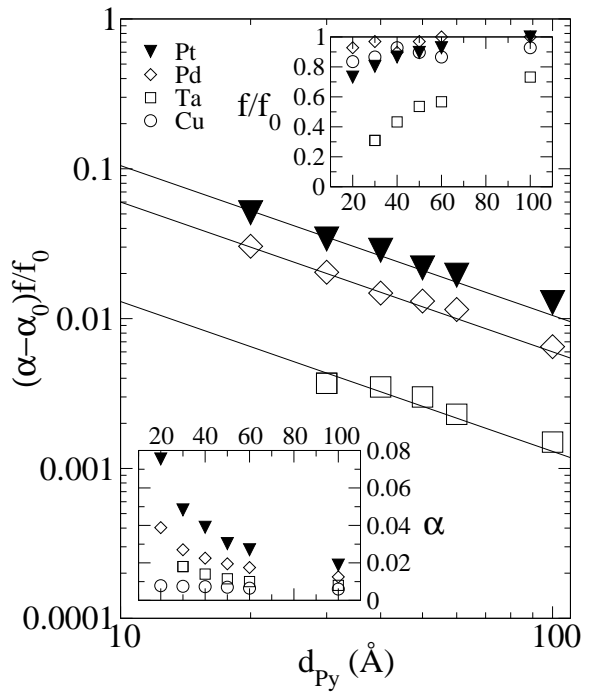}
\caption{The lines show our theoretical result (9)
with $\kappa=$1.0, 0.6, and 0.1; the data points are derived
from the measurements [12] shown in two insets.
Insets: Measured Gilbert damping constant $\alpha$ (lower inset)
and the relative atomic magnetization $f/f_{0}$ (upper inset) in permalloy film of varied thickness $d_{\text{Py}}$ in a trilayer structure \textit{N}-Py-\textit{N}.}
\label{f2}
\end{figure}

Let us now estimate the damping coefficient $\alpha ^{\prime }$ for thin
films of permalloy (Ni$_{80}$Fe$_{20},$ Py), a magnetically very soft
material of great technological importance. Mizukami \textit{et al.} \cite
{Mizukami:jjap01} measured the ferromagnetic resonance linewidth of
\textit{N}-Py-\textit{N}
sandwiches and discovered systematic trends in the damping
parameter as a function of Py layer thickness $d$ for different normal
metals. The spin polarization of electrons emitted by Py has been measured
to be $p\approx 0.4$ in point contacts \cite{Soulen:sc98}, the magnetization
per atom is $f\approx 1.2$, and Land\'{e} factor--$g_{L}\approx 2.1$ \cite
{Mizukami:jjap01}. The interface conductance of metallic interfaces with Fe
or Co is of the order of 10$^{15}$~$\Omega ^{-1}$m$^{-2}$, with significant
but not drastic dependences on interface morphology or material combination 
\cite{Xia:prb01}. This corresponds to roughly one conducting channel per
interface atom. Assuming the Fermi surface of the normal metal is isotropic,
we arrive at the estimate $\alpha ^{\prime }\approx 1.1/d(\text{\AA })$. The
factor $1/d$ does not reflect an intrinsic effect; a reduced total
magnetization is simply more sensitive to a given spin-current loss at the
interface. Comparing with the intrinsic $\alpha _{0}\approx 0.006$ of
permalloy \cite{Patton:jap75,Mizukami:jjap01} the spin-current induced
damping becomes important for ferromagnetic layers with thickness $d<100$~\AA.
We can refine the estimate by including the significant film-thickness
dependence of the magnetization measured by the same group \cite
{Mizukami:jjap01}. We, therefore, improve our above estimate as 
\begin{equation}
\alpha ^{\prime }(d)\approx \kappa \times \frac{1.1}{d(\text{\AA })}\times 
\frac{f_{0}}{f(d)}\,,  \label{es}
\end{equation}
where $f_{0}$ and $f(d)$ are the atomic magnetization of the permalloy bulk
and films. $\kappa $ is an adjustable parameter representing the number of
scattering channels in units of one channel per interface atom, which should
be of the order of unity.

The experimental results for the damping factor $\alpha $ and the relative
magnetization $f/f_{0}$ for \textit{N}-Py-\textit{N} sandwiches with \textit{
N}=Pt, Pd, Ta, and Cu are shown in the insets of Fig.~\ref{f2}. Our estimate
(\ref{es}) appears to well explain the dependence of $\alpha $ on the
permalloy film thickness $d$ (see Fig.~\ref{f2}) for reasonable values of $
\kappa $. First-principles calculations are called for to test these values.

The lack of a significant thickness dependence of damping parameter of the
Cu-Py system requires additional attention. An opaque interface might be an
explanation, but it appears more likely that due to long spin-flip
relaxation times in Cu, the 5 nm thick buffer layers in \cite{Mizukami:jjap01}
do not provide the ideal sink for the injected spins as assumed above. This
means that a nonequilibrium spin accumulation on Cu opposes the pumped spin
current and nullifies the additional damping when
$h/\tau_{\text{sf}}\delta$
is comparable or smaller than the conductance $g$. For 5 nm Cu buffers,
$g\delta/h\sim 10^{13}$~s$^{-1}$,
whereas $1/\tau _{\text{sf}}\sim 10^{12}$~s$^{-1}$ \cite{Meservey:prl78}.
It follows that Cu is indeed a poor sink for
the injected spins and the Gilbert damping constant is not enhanced. On the
other hand, Pt, Ta, and Pd are considerably heavier than Cu and, since $
1/\tau _{\text{sf}}$ scales as $Z^{4}$ \cite{Abrikosov:zetf62}, where $Z$ is
the atomic number, have much larger spin-relaxation rates and our arguments
hold.

A physical picture of the effect of magnetization precession in layered
systems has been proposed earlier by Hurdequint \textit{et al.}
\cite{Monod:prl72} in order to explain
ferromagnetic and conduction electron spin
resonance experiments. These authors realized that the precessing
magnetization is a source of a nonequilibrium spin accumulation which
diffuses out of the \textit{N-F} interfaces into the adjacent normal metal
layers where it can dissipate by spin-flip processes. Enhanced Gilbert
damping in thin ferromagnetic films in contact with normal metal has also
been discussed by Berger \cite{Berger:prb96} for a ballistic \textit{N-F}
interface in a spin-valve configuration. His expression for the damping
coefficient [Eq.~(20) in Ref.~\cite{Berger:prb96}] scales like ours as
a function of layer thickness, but differs as a function of material
parameters. \textit{E.g.}, in contrast to our result, Berger's expression
does not vanish with vanishing exchange splitting.

In conclusion, we demonstrated that the Gilbert damping constant is enhanced
in thin magnetic films with normal metal buffer layers by a spin-pump effect
through the \textit{N-F} contact. The damping is significant for transition
metal films thinner than about 10 nm. Recent experiments on permalloy films 
\cite{Mizukami:jjap01} are well explained.

We are grateful to
L.~Berger, B.~I.~Halperin, B.~Heinrich, D.~Huertas-Hernando, and Yu.~V.~Nazarov
for stimulating discussions. This work was supported in part by
the DARPA Award No. MDA 972-01-1-0024,
the Norwegian Research Council,
the NEDO International Joint Research Grant Program
\textquotedblleft Nano-magnetoelectronics\textquotedblright , NSF Grant No. DMR
99-81283, and the Schlumberger Foundation.

\textit{Note added.}---After submission of this paper, we became aware of new
exciting experimental results \cite{Heinrich:prl01} on enhanced Gilbert
damping in ultrathin iron films which support the
spin-pumping mechanism proposed in this Letter \cite{Tserkovnyak:prep}.

\end{document}